\documentclass[11pt]{oxarticle}
\pdfoutput=1

\usepackage{graphicx, float, array, xspace, amscd, amsmath, amsthm, amssymb, latexsym, relsize, bbm}
\usepackage[letterpaper, left=0.78in, right=0.78in, top=0.68in, bottom=0.72in]{geometry}
\usepackage[bbgreekl]{mathbbol}
\usepackage{amsfonts, tikz-cd}
\usepackage[T1]{fontenc}

\usepackage{tikz}
\usetikzlibrary{backgrounds}
\usepackage[framemethod=tikz]{mdframed}
\AtBeginEnvironment{mdframed}{%
\tikzset{every picture/.style={}}%
}
\mdfsetup{roundcorner=.5ex}
{\begin{mdframed}[backgroundcolor=white]}%
{\end{mdframed}}

\DeclareSymbolFontAlphabet{\mathbb}{AMSb}
\DeclareSymbolFontAlphabet{\mathbbl}{bbold}

\usepackage[sort&compress, comma, square, numbers]{natbib}
\usepackage[all,cmtip]{xy}
\usepackage{color}
\definecolor{MyDarkBlue}{rgb}{0.15,0.25,0.45}
\usepackage{braket}      
\usepackage[linktocpage=true]{hyperref}
\hypersetup{colorlinks=true, citecolor=blue, linkcolor=blue, urlcolor=blue, pdfauthor={}, pdftitle={},breaklinks=true,pageanchor=false, hypertexnames=false}

\numberwithin{equation}{section}

\newcommand{\sh}{{\sharp}}

\newcommand{\LC}{\text{\tiny LC}}

\renewcommand{\H}{{\text{\tiny H}}}

\newcommand{\Ch}{\text{\tiny CH}}

\newcommand{\rb}{{\overline{ r}}}

\newcommand{\Ob}{{\overline{ \Omega}}}

\newcommand{\w}{{\,\wedge\,}}
\newcommand{\wt}{\widetilde}
\newcommand{\wh}{\widehat}

\newcommand{\fD}{{\mathfrak{D}}}

\newcommand{\half}{\frac{1}{2}}

\newcommand{\ab}{{\overline\alpha}}
\newcommand{\bb}{{\overline\beta}}

\newcommand{\A}{\cA}
\newcommand{\M}{\ccM}

\renewcommand{\aa}{\mathfrak{a}}

\renewcommand{\a}{\alpha}

\renewcommand{\d}{\delta}\newcommand{\D}{\Delta}
\newcommand{\e}{\epsilon}\newcommand{\ve}{\varepsilon}

\newcommand{\Th}{\Theta}

\renewcommand{\L}{\Lambda}
\newcommand{\m}{\mu}
\newcommand{\n}{\nu}
\newcommand{\x}{\xi}

\renewcommand{\r}{\rho}
\newcommand{\s}{\sigma}
\renewcommand{\t}{\tau}

\renewcommand{\o}{\omega}\renewcommand{\O}{\Omega}

\DeclareFontFamily{OT1}{pzc}{}
\DeclareFontShape{OT1}{pzc}{m}{it}{<-> s * [1.200] pzcmi7t}{}
\DeclareMathAlphabet{\mathpzc}{OT1}{pzc}{m}{it}

\newcommand{\cA}{\mathcal{A}}

\newcommand{\cE}{\mathcal{E}}

\newcommand{\ccM}{\mathpzc M}

\newcommand{\cO}{\mathcal{O}}

\newcommand{\ccT}{\mathpzc T}
\newcommand{\ccU}{\mathpzc U}

\newcommand{\ccZ}{\mathpzc Z}

\DeclareFontFamily{U}{bbold}{}
\DeclareFontShape{U}{bbold}{m}{n}
 {  <-5.5> s*[1.05] bbold5
    <5.5-6.5> s*[1.05] bbold6
    <6.5-7.5> s*[1.05] bbold7
    <7.5-8.5> s*[1.05] bbold8
    <8.5-9.5> s*[1.05] bbold9
    <9.5-11.5> s*[1.05] bbold10
    <11.5-16> s*[1.05] bbold12
    <16-> s*[1.05] bbold17
 }{}

\newcommand{\IA}{\mathbbl{A}}
\newcommand{\IB}{\mathbbl{B}}

\newcommand{\Id}{\mathbbl{d}}

\newcommand{\IF}{\mathbbl{F}}

\newcommand{\IH}{\mathbbl{H}}

\renewcommand{\IJ}{\mathbbl{J}}

\newcommand{\IQ}{\mathbbl{Q}}
\newcommand{\IR}{\mathbbl{R}}

\newcommand{\IX}{\mathbbl{X}}

\newcommand{\IDb}{{\overline{\mathbbl{D}}}}

\newcommand{\ITheta}{\mathbbl{\Theta}}

\newcommand{\Iomega}{\mathbbl{\hskip1pt\bbomega}}

\makeatletter
\newcommand{\Idelcore}[2]{%
  \ooalign{%
    $\m@th#1\partial$\cr
    \hidewidth\raisebox{0.02ex}{\vrule height .55em depth .08em width .048em}\hidewidth\cr
  }%
}
\newcommand{\Idel}{{\mathord{\mathpalette\Idelcore\relax}}}
\newcommand{\Idelb}{{\overline{\Idel}}}
\makeatother

\newcommand{\Deth}{\text{\DH}\hskip1pt}
\newcommand{\Dethsharp}{\Deth^\sharp}

\font\eightrm=cmr8 at 8pt

\newcommand{\beq}{\begin{equation}}
\newcommand{\eeq}{\end{equation}}
\newcommand{\beqnn}{\begin{equation*}}
\newcommand{\eeqnn}{\end{equation*}}
\newcommand{\bea}{\begin{eqnarray}}
\newcommand{\eea}{\end{eqnarray}}
\newcommand{\bean}{\begin{eqnarray*}}
\newcommand{\eean}{\end{eqnarray*}}

\newcommand{\norm}[1]{\left\| #1\right\|}

\newcommand{\ii}{\text{i}}

\newcommand{\place}[3]{\vbox to0pt{\kern-\parskip\kern-7pt
 \kern-#2truein\hbox{\kern#1truein #3}
 \vss}\nointerlineskip}

\DeclareFontFamily{U}{wncy}{}
\DeclareFontShape{U}{wncy}{m}{n}{<->wncyr10}{}
\DeclareSymbolFont{mcy}{U}{wncy}{m}{n}
\DeclareMathSymbol{\sha}{\mathord}{mcy}{"58}

\newcommand{\del}{{\partial}}
\newcommand{\delb}{{\overline{\partial}}}

\newcommand{\lb}{{\overline\lambda}}
\newcommand{\nb}{{\overline\n}}
\newcommand{\mb}{{\overline\m}}

\newcommand{\Dbar}{{\overline D}}

\renewcommand{\aa}{\mathfrak{a}}

\newcommand{\dd}{{\text{d}}}

\newcommand{\K}{K\"ahler\xspace}

\renewcommand{\H}{\text{H}}

\newcommand{\tr}{\text{Tr}\hskip2pt}

\newcommand{\tb}{{\overline{\tau}}}

\newcommand{\ap}{{\a^{\prime}\,}}

\renewcommand{\rb}{{\overline{\rho}}}
\renewcommand{\=}{\;=\;}

\makeatletter
\g@addto@macro\bfseries{\boldmath}
\makeatother

\newcommand{\citeM}{\cite{Candelas:2016usb}\xspace}

\newcommand{\citeUG}{\cite{Candelas:2018lib}\xspace}

\newcommand{\citeAPmetric}{\cite{McOrist:2026cys}\xspace}
\newcommand{\citeAP}{\cite{McOrist:2025zwf}\xspace}

\proofmodefalse

\usepackage{microtype}
\allowdisplaybreaks[2]
\hypersetup{pdftitle={Universal geometry as an organising principle for heterotic moduli}}

\begin{document}
\pagestyle{plain}
\pagenumbering{arabic}

\begingroup
\definecolor{UGblue}{RGB}{38,82,116}
\definecolor{UGbluefill}{RGB}{229,239,247}
\definecolor{UGorange}{RGB}{177,96,32}
\definecolor{UGorangefill}{RGB}{248,234,219}
\definecolor{UGgreen}{RGB}{50,111,88}
\definecolor{UGgray}{RGB}{88,96,103}
\definecolor{UGgrayfill}{RGB}{241,243,244}
\definecolor{UGink}{RGB}{31,37,42}

\tikzset{
  >=Latex,
  every node/.style={text=UGink, font=\small},
  outer family/.style={draw=UGgray!78, fill=UGgrayfill!45,
    dash pattern=on 4pt off 2.6pt, rounded corners=5pt, line width=.75pt},
  compact family/.style={draw=UGblue!72, fill=UGbluefill!42,
    rounded corners=4pt, line width=.8pt},
  fibre/.style={draw=UGblue, fill=UGbluefill, line width=1pt},
  gauge/.style={draw=UGorange, line width=1.15pt},
  lift/.style={draw=UGgreen, line width=1.05pt,
    -{Latex[length=2.2mm,width=1.45mm]}},
  projection/.style={draw=UGgray!82, line width=.7pt,
    -{Latex[length=2mm,width=1.35mm]}},
  leader/.style={draw=UGgray!72, line width=.55pt},
  hierarchy arrow/.style={line width=.68pt, line cap=round,
    -{Latex[length=1.85mm,width=1.25mm]}},
  base/.style={draw=UGgray, fill=UGgrayfill, line width=.8pt}
}

\begin{center}

{\Large\bfseries Universal geometry as an organising principle for heterotic moduli\par}
\vspace{0.35em}
{\normalsize Jock McOrist\footnote{\href{mailto:jmcorist@une.edu.au}{\texttt{jmcorist@une.edu.au}}}\par}
\vspace{0.25em}
{\small School of Science and Technology\par}
{\small University of New England\par}
{\small  Armidale, NSW, Australia\par}
\vspace{0.55em}

\begin{minipage}{0.91\textwidth}
\small
\textbf{Abstract:}
A family of heterotic compactifications carries more structure than a collection of solutions parametrised by moduli. Once the compactification data are fibred over moduli space, deformations become components of universal curvatures. This note reviews that organisation and explains how it incorporates the $\ap^2$ supersymmetry corrections.

\end{minipage}
\end{center}
\vspace{-0.25em}

\section{The problem hidden in a family of vacua}

Consider a perturbative $d=4$, $N=1$ heterotic vacuum on a complex threefold $X$ with $c_1(X)=0$, Hermitian form $\o$, holomorphic volume form $\O$ and gauge connection $\A$ on $E\to X$.  Let $\Th=\Th^\H$ be the  Hull connection on $\ccT_X$, with $\Th^\H_m=\Th^\LC_m+\half H_m$, and write its curvature as $R=R^\H$ \cite{Hull:1986kz,Strominger:1986uh,Bergshoeff:1989de}.  It was shown in \citeAP that the $\ap^2$ corrections can be  accounted for by introducing hats
\beq\label{eq:hatted-field-redefinition}
 \widehat g_{\m\nb}\=g_{\m\nb}+\frac{\ap^2}{4}\cE^\H_{\m\nb}~,
 \qquad
 \widehat\o\=\o+\frac{\ii\ap^2}{4}\cE^\H_{\m\nb}\dd x^{\m\nb}~,
 \qquad
 \widehat\Phi\=\Phi+\frac{3\ap^2}{32}h~,
\eeq
where $h=\tr|F|^2-\tr|R^\H|^2$ and $\cE^\H_{mn}=\tr(F_{mp}F_n{}^p)-\tr(R^\H_{mp}R^{\H}{}_n{}^p)$. 
Supersymmetry and anomaly cancellation require 
\beq\label{eq:susyHS}
N_J\=0~,\qquad F^{0,2}\=0~,\qquad H\=\dd^c\wh \o~,\qquad
\dd H\= \frac{\ap}{4}\big(\tr F\w F-\tr R^\H\w R^\H\big)~,
\eeq
together with
\beq\label{eq:hatted-D-terms}
 \log\norm{\O}_{\widehat g}=-2\widehat\Phi+\cO(\ap^3)~,
 \qquad
 \dd\bigl(e^{-2\widehat\Phi}\widehat\o^2\bigr)=\cO(\ap^3)~,
 \qquad
 \ap\,F\wedge\widehat\o^2=\cO(\ap^3)~.
\eeq
Remarkably, the equations take the form of the Hull--Strominger system when written in terms of the hatted fields.  The $\ap^2$ corrections have not disappeared but are encoded in the field redefinitions. In particular, \citeAP found a new $\ap^2$ term in the graviton equation of motion, most easily derived using a lemma of \cite{Bergshoeff:1989de}. Together with results from Hermitian geometry, equations \eqref{eq:susyHS}--\eqref{eq:hatted-D-terms} imply all the equations of motion, including this term. The hatted fields therefore package the $\ap^2$ corrections while preserving the leading-order form of the equations, making this basis a natural starting point for the higher-order expansion.\footnote{A formulation using another connection, such as the Chern connection, is related to the Hull choice by an $\ap$-dependent field redefinition. In those variables the F-terms acquire explicit $\ap^2$ corrections. If the redefinition depends on the local $B$-field potential, the resulting metric need not be globally defined \cite{Melnikov:2014ywa}.}
 
The proof also uses the fact that the Hull connection is composite: it is built from $g$ and $H$. Both $(R^\H)^{0,2}$ and $ R^\H\w\o^2$ acquire source terms controlled by $\dd H$: this means it is not an instanton, and this is crucial to the result in \citeAP, in which supersymmetry and the Bianchi identity imply the equations of motion. Finally, one can apply the spinorial supersymmetry variations twice and recover the equations of motion provided the Bianchi identities are imposed. This circle of ideas gives several intricate and non-trivial consistency checks of the theory.

The Hull--Strominger system is a low-order truncation of the string theory sigma model \cite{Hull:1986xn,Ross:1986ra,Hull:1987pc,Metsaev:1986yb,Cai:1986sa,Metsaev:1987zx,Gross:1986mw,Foakes:1988wy,Chemissany:2007he}, and is not protected to all orders. For example, in the Calabi--Yau sigma model arising from the standard embedding in the heterotic string, the four-loop beta function gives a $\ap^3\zeta(3)$ correction to the Hull--Strominger equations \cite{Grisaru:1986dk,Gross:1986iv,Melnikov:2014ywa}, resulting in a metric in the same \K class that is no longer Ricci-flat.  The corrections \eqref{eq:hatted-field-redefinition} are analogous to this. 

The value of this simple form is exemplified in the moduli problem, where complex-structure, Hermitian, bundle and $B$-field deformations are coupled rather than forming independent sectors.  The familiar labels ``complex-structure modulus'', ``bundle modulus'' and ``\K modulus'' are  useful only at special loci or at zeroth order in $\ap$.  At first order in $\ap$, the deformation problem is organised
by the reduced double extension $Q$ \cite{McOrist:2021dnd}, which builds on earlier work of \cite{delaOssa:2014cia,Anderson:2014xha}.  As a smooth
bundle,
\beq\label{eq:Qfibre}
Q\simeq
(\ccT_X^{1,0})^*\oplus{\rm End}\,E\oplus\ccT_X^{1,0}~.
\eeq
We write its sections as $\mathpzc Y=(\mathpzc Z,\mathfrak a,\Delta)$.  The variation of the composite Hull connection is already expressed in terms of these fields.  The F-term equations and the Bianchi identity give
\beq
\Dbar^2=\cO(\ap^2)~,
\eeq
so $\Dbar\mathpzc Y=0$ describes the F-term deformations at first order in $\ap$.  At the same order, $\Dbar^\dag\mathpzc Y=0$ selects the D-term representatives \cite{McOrist:2021dnd} and there is a close relation to a generalisation of Hermitian Yang-Mills on the extension bundle \cite{McOrist:2024zdz}.  The F-term equations themselves extend to order $\ap^2$, as shown below, but the corresponding reduced operator and adjoint have not yet been constructed.

The extension picture \eqref{eq:Qfibre} organises first-order deformations of one solution, but does not say how the deformation spaces at neighbouring vacua are related. This is the role of universal geometry. Start with a local family of solutions over $\M$, a local smooth patch of the moduli space. Assume $X_y$ form a smooth fibration $\pi:\IX\to\M$ and that the $E_y\to X_y$ assemble into a vector bundle $\ccU\to\IX$:
\begin{center}
\begin{minipage}{0.94\textwidth}
\centering
\par\vspace{0.2em}
\begin{tikzcd}[column sep=3.7em,row sep=0.95em]
E_y \arrow[r,hook] \arrow[d] & \ccU \arrow[d] \\
X_y \arrow[r,hook] \arrow[d] & \IX \arrow[d,"\pi"] \\
\{y\} \arrow[r,hook] & \M
\end{tikzcd}
\end{minipage}
\end{center}
On $\IX$, we choose a horizontal distribution and assume it is integrable. Said differently, there is Ehresmann connection, introduced below, which has vanishing curvature $S=0$. This gives a splitting into vertical and horizontal components and makes the family locally a product \cite{Candelas:2018lib,McOrist:2024glz}. Using the splitting we introduce on the total space an almost complex structure  $\IJ = J \oplus J^\sh$ and almost hermitian form $\wh \Iomega = \wh\o \oplus \o^\sh$, where $J^\sh$ and $\o^\sh$ are the complex structure and \K form on $\M$ respectively. Let $\IH$ be a three-form and $\IA$ a connection for $\ccU$ with field strength $\IF$.  Denote $\iota_y:X_y\hookrightarrow\IX$ to be the inclusion map.   We require
\beq
\IJ|_{TX_y} \= J_y~,\qquad
\iota_y^*\IA=A_y~,\qquad
\iota_y^*\wh\Iomega=\wh\o_y~,\qquad
\iota_y^*\IH=H_y~.
\eeq
These conditions fix the vertical components of the universal tensor and  satisfy \eqref{eq:susyHS}--\eqref{eq:hatted-D-terms} by assumption. Components with legs along $\M$ describe how the fields on nearby fibres are identified and are not unique.  These are related to connections and discuss in the next section. The collection of fields on $\IX$ and its bundle $\ccU$ are referred to as universal heterotic geometry.

Forms on $\IX$ carry two gradings. Their base/fibre degree, called tangibility, counts the legs along $\M$ and along the fibre. Their holomorphic type is defined by the complex structures of $\M$ and $X_y$. The first grading separates background equations, first-order deformations and moduli-space curvatures; the second selects the holomorphic sector relevant to the F-terms. In the universal approach, tensor equations on $\IX$ unify the F-terms and their deformations in a compact way.

This note reviews how universal curvatures encode heterotic deformations and then applies this framework to the $\ap^2$ results of \citeAP. In the hatted variables, the corrected fibre equations and their mixed components combine into universal tensor equations through order $\ap^2$. We demonstrate that the lift to the universal geometry is consistent at $\ap^2$ with the Hull curvature $\IR^\H$. 

\vskip1cm

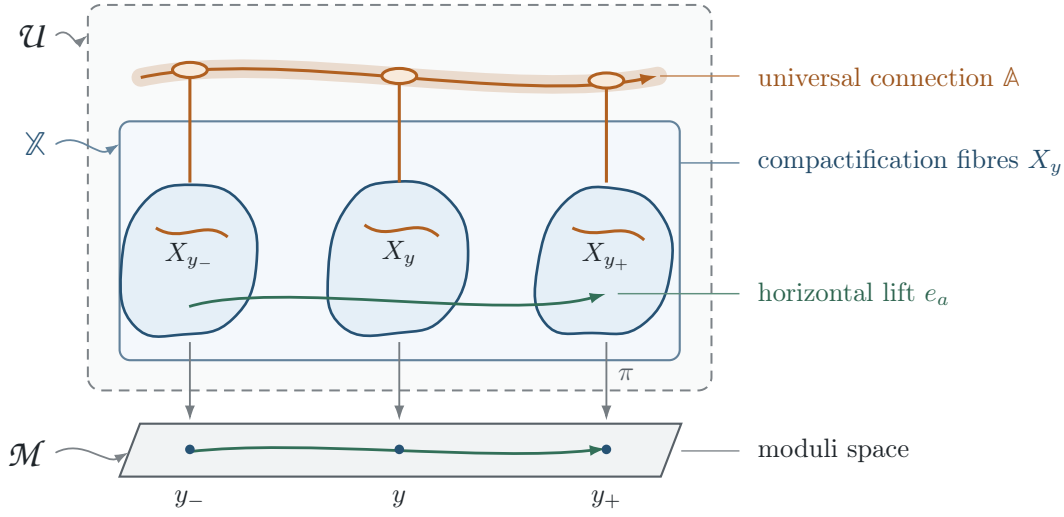
\begin{tikzpicture}[x=1cm,y=1cm]
  \useasboundingbox (-0.10,0.05) rectangle (13.90,7.25);

  \path[outer family] (1.30,1.72) rectangle (9.55,6.82);
  \path[compact family] (1.72,2.12) rectangle (9.13,5.28);

  \path[base]
    (1.72,0.58) -- (8.88,0.58) -- (9.15,1.28) -- (1.99,1.28) -- cycle;

  \draw[lift] (2.64,0.91) .. controls (4.10,1.08) and (6.62,0.78) .. (8.16,0.95);
  \foreach \x/\lab in {2.65/{y_-},5.42/{y},8.16/{y_+}}{
    \fill[UGblue] (\x,0.94) circle (1.8pt);
    \node[anchor=north] at (\x,0.53) {$\lab$};
  }

  \begin{scope}[shift={(2.65,3.45)}]
    \path[fibre] plot[smooth cycle,tension=.82] coordinates
      {(-.73,-.88) (-.91,-.13) (-.63,.73) (.04,.99)
(.70,.72) (.88,-.06) (.57,-.78) (-.09,-.99)};
    \node at (0,.05) {$X_{y_-}$};
    \draw[gauge] (-.45,.42) .. controls (-.10,.20) and (.19,.55) .. (.50,.30);
  \end{scope}
  \begin{scope}[shift={(5.42,3.48)}]
    \path[fibre] plot[smooth cycle,tension=.82] coordinates
      {(-.76,-.91) (-.94,-.19) (-.68,.69) (-.08,1.00)
(.64,.78) (.91,.02) (.62,-.76) (-.02,-1.00)};
    \node at (0,.05) {$X_y$};
    \draw[gauge] (-.46,.43) .. controls (-.12,.18) and (.20,.57) .. (.52,.28);
  \end{scope}
  \begin{scope}[shift={(8.16,3.44)}]
    \path[fibre] plot[smooth cycle,tension=.82] coordinates
      {(-.81,-.83) (-.91,-.08) (-.55,.77) (.13,.95)
(.72,.61) (.85,-.14) (.49,-.84) (-.19,-.96)};
    \node at (0,.05) {$X_{y_+}$};
    \draw[gauge] (-.45,.43) .. controls (-.08,.25) and (.20,.54) .. (.50,.27);
  \end{scope}

  \foreach \x in {2.65,5.42,8.16}{
    \draw[projection] (\x,2.36) -- (\x,1.32);
  }
  \node[anchor=west, text=UGgray, inner sep=0pt]
    at (8.31,1.94) {$\pi$};

  \draw[gauge, opacity=.20, line width=7pt, line cap=round]
    (2.00,5.86) .. controls (3.80,6.25) and (6.96,5.50) .. (8.84,5.88);
  \draw[gauge, -{Latex[length=2.3mm,width=1.55mm]}]
    (2.00,5.86) .. controls (3.80,6.25) and (6.96,5.50) .. (8.84,5.88);

  \foreach \x/\y in {2.65/5.96,5.42/5.88,8.16/5.82}{
    \draw[gauge] (\x,4.47) -- (\x,\y-0.09);
    \draw[gauge, fill=UGorangefill] (\x,\y) ellipse (.23 and .10);
  }

  \draw[lift]
    (2.64,2.83) .. controls (3.78,3.18) and (6.86,2.64) .. (8.16,3.00);

  \node[anchor=east, font=\bfseries\large] (Ulab) at (0.88,6.42) {$\ccU$};
  \draw[hierarchy arrow, draw=UGgray!78]
    (Ulab.east) .. controls +(0.16,0.13) and +(-0.16,-0.13) .. (1.30,6.42);

  \node[anchor=east, font=\bfseries\large, text=UGblue] (Xlab) at (0.88,4.98) {$\IX$};
  \draw[hierarchy arrow, draw=UGblue!70]
    (Xlab.east) .. controls +(0.30,0.18) and +(-0.30,-0.18) .. (1.72,4.98);

  \node[anchor=east, font=\bfseries\large] (Mlab) at (0.88,0.90) {$\M$};
  \draw[hierarchy arrow, draw=UGgray!78]
    (Mlab.east) .. controls +(0.35,0.18) and +(-0.30,-0.16) .. (1.84,0.90);

  \draw[leader, draw=UGorange!78] (8.87,5.88) -- (9.92,5.88);
  \node[anchor=west, text=UGorange!88!black] at (10.03,5.88)
    {universal connection $\IA$};

  \draw[leader, draw=UGblue!70] (9.13,4.73) -- (9.92,4.73);
  \node[anchor=west, text=UGblue!90!black] at (10.03,4.73)
    {compactification fibres $X_y$};

  \draw[leader, draw=UGgreen!78] (8.31,3.00) -- (9.92,3.00);
  \node[anchor=west, text=UGgreen!90!black] at (10.03,3.00)
    {horizontal lift $e_a$};

  \draw[leader] (9.15,0.92) -- (9.92,0.92);
  \node[anchor=west] at (10.03,0.92) {moduli space};

\end{tikzpicture}
\endgroup

\vskip1cm

\section{Universal connections and holomorphic gauge}
To compare solutions on nearby fibres we need to introduce connections associated to gauge symmetries. First, consider parameter-dependent diffeomorphisms $x^m\mapsto \widetilde x^m(x,y)$. This requires an Ehresmann connection $c_a{}^m$ and the adapted basis
\beq\label{eq:ebasis-letter}
e^m\=\dd x^m+c_a{}^m\dd y^a~,
\qquad
 e_a\=\del_a-c_a{}^m\del_m~.
\eeq
The connection $c_a{}^m$ specifies the horizontal subspace. It has a curvature $S$, which vanishes as we have assumed the product structure is integrable. This means that nearby fibres can be identified consistently: transporting a point around a loop in moduli space  produces no diffeomorphism of the fibre. The family does not fix $c_a{}^m$ uniquely. Different choices are related by parameter-dependent diffeomorphisms. In holomorphic coordinates,
$$
 \Delta_{\alpha\nb}{}^\mu\=\!-\del_{\nb}c_\alpha{}^\mu~.
$$
Under $c_\alpha{}^\mu \to c_\a{}^\m - \ve_\a{}^\m$, $\D_\a{}^\m \to \D_\a{}^\m + \delb \ve_\a{}^\m$. Thus changing the horizontal lift changes the representative of the Kodaira--Spencer class. 

Similarly, we allow gauge transformations to depend on both $x$ and $y$ and this requires an extended connection
\beq\label{eq:universal-connection}
\IA\=A+A_a^{\sharp}\dd y^a~, \qquad\qquad \IF\=\Id\IA+\IA^2~.
\eeq
Here $A$ is the gauge connection on each fibre. The component $A_a^\sharp$ identifies the gauge fields on nearby fibres. It is not a new physical field and changing it gives the small gauge transformation described below. In the basis \eqref{eq:ebasis-letter}, its curvature decomposes as
\beq\label{eq:curvature-decomp}
\IF\=\half F_{mn}e^me^n+\dd y^a\,\fD_a A
      +\half\dd y^a\dd y^b\,\IF_{ab}~.
\eeq
These terms have tangibility $[0,2]$, $[1,1]$ and $[2,0]$ respectively.  The middle component, the mixed-tangibility piece $[1,1]$, contains the covariant deformation $\fD_aA$.  The following identities
\beq\label{eq:bianchi-components}
\dd_A(\fD_aA)\= \fD_a F~,
\qquad
[\fD_a,\fD_b]A \=\! -\dd_A\IF_{ab}~,
\eeq
proved by hand in \citeM are also components of the universal Bianchi identity $\Id_\IA\IF=0$.

These curvature components also give an invariant meaning to gauge fixing.  There are two notions of gauge transformation in a background-field expansion.  Background transformations act simultaneously on the fields and their moduli-space connections.  Small transformations act on the fluctuation representatives and must be quotiented out to obtain physical fields.  For deformations of $A$, changing the horizontal component by $A_a^{\sharp}\mapsto A_a^{\sharp}-\phi_a$ gives
\beq\label{eq:small-gauge}
\fD_a \A\longmapsto \fD_a\A+\delb_\A\phi_a~,
\eeq
which is exactly a small gauge transformation.  A gauge slice on fluctuations is therefore equivalent to a choice of horizontal lift on the universal bundle \cite{McOrist:2019mxh}.

For the universal gauge bundle, the equation
\beq\label{eq:holomorphic-universal}
\IF^{0,2}\=0~,
\eeq
contains three statements. Its vertical part is the fibre equation $F^{0,2}=0$. Its mixed part sets the anti-holomorphic variation $\IF_{\ab\mb} = \fD_\ab \A_\mb=0$; this is holomorphic gauge and partly selects a horizontal lift. The component $\IF_{\ab\bb} = 0$ follows by a consistency calculation \cite{McOrist:2024glz}; it says that the connection along $\M$ is holomorphic. Thus gauge fixing is closely tied to the holomorphic structure of the universal bundle. The Ehresmann connection carries the diffeomorphism small gauge choice while horizontal conditions for $\IB$ likewise encode the gerbe gauge choice. As mentioned above, the hermitian form $\wh \Iomega$ and $\IJ$ are block diagonal in the adapted basis \eqref{eq:ebasis-letter}. A tangent direction on $\M$ has an intrinsic holomorphic type, but the corresponding variation of a field representative is defined only after choosing the universal connections that identify neighbouring fibres.  Holomorphic gauge is a necessary part of the statement that these choices define holomorphic structures on the family.

The curvature $\IF_{ab}$ then has an immediate interpretation: it measures the failure of the chosen horizontal lifts and hence of gauge-fixed deformations to commute.  This is why second-order deformation theory necessarily sees connections on $\M$: commutators of covariant derivatives are curvatures of the universal geometry.

In contrast with $\IA$, $\ITheta^\H$ is not an independent gauge field: it is the lift of the family of connections $\Th^\H(g_y,H_y)$.  Consequently its mixed-tangibility curvature is the physical spin-connection deformation,
\beq\label{eq:mixed-Hull-preview}
 (\IR^\H)_{am}e^m \= \fD_a\Th^\H~.
\eeq
For the Hull connection, the analogue of \eqref{eq:holomorphic-universal} is subtle.  There is no independent horizontal component of $\ITheta^\H$ to choose: the connection is composite, so its  curvature must be computed from the universal lifts of $g$ and $H$.  It is this curvature that enters the universal F-terms and Green--Schwarz Bianchi identity.

\section{Universal F-terms and the Hull curvature}
Given the horizontal distribution is integrable, we find the F-term and anomaly equations hold on the total space
\beq\label{eq:universal-system}
N_\IJ\=0~, \quad \IF^{0,2}+  \cO(\ap^3) \=0~, \quad \IH\=\Id^c\wh \Iomega + \cO(\ap^3)~, \quad \Id\IH\=\frac{\ap}{4}\big(\tr\IF\w\IF-\tr\IR^\H\w \IR^\H\big)~.
\eeq
Its vertical components hold because every fibre is a solution. Its mixed components are checked below through order $\ap^2$. Pure-base components require second-order deformation theory and are established only through order $\ap$, under the assumptions of Section 4. 

The first two equations have different roles. To prove $N_\IJ=0$, decompose into tangibility. With the curvature $S=0$, three conditions are required: $N_J = 0$ fibrewise; the usual Kodaira--Spencer condition $\delb \D_\a = 0$; and $N_{J^\sh} = 0$. All of these conditions hold in virtue of $N=1$ supersymmetry.  By contrast, as discussed above, the vertical part of $\IF^{0,2}=0$ is holomorphy of $E$ while the remaining components amount to holomorphic gauge and a consistency condition for the moduli space connection. Together $N_\IJ = 0$, $\IF^{0,2} = 0$ make $\IX$ complex and $\ccU$ holomorphic. 

To prove the third equation of \eqref{eq:universal-system}, we again decompose by tangibility. The fibre component holds by $\ap^2$ supersymmetry relations in the hatted basis, while the component with one moduli leg is the linearised F-term equation. The components with two moduli legs vanish by the product structure and the block form of $\wh\Iomega$, see \eqref{eq:flatness-letter} below. The component with three moduli legs vanishes because $\M$ is \K. For the last equation, the fibre component is the Bianchi identity. Its mixed component is its linearisation, while components with two moduli legs give the second-order relations derived directly in \citeM.

The component $\IR_{am}^\H$ in \eqref{eq:mixed-Hull-preview} gives a non-trivial consistency check of this construction.  At zeroth order\footnote{Since the Green--Schwarz identity carries an overall factor of $\ap$, the $\cO(\ap)$ part of $\IR^\H$ contributes at order $\ap^2$ to the deformation equations.} it reproduces the direct variation of $g$ and $H$ found in \cite{McOrist:2025sdy}; at first order it reproduces the results in \citeAPmetric.  Subsection~\ref{s:mixedsub} computes the $(\IR^\H_{\alpha\mb})^\nu{}_{\rho}$ component, while Subsection~\ref{s:app-off-diagonal-Hull-blocks} records the $(\IR^\H_{\alpha\mb})^{\nb}{}_{\rho}$ and $(\IR^\H_{\alpha\mb})^\nu{}_{\rb}$ components.  Together these give the components of $\fD_\alpha\Theta^{\H\,0,1}$ used in the $\ap^2$ correction to the moduli-space metric, using the formalism of \citeUG.  The comparison is nontrivial because the two derivations organise the same data differently: the universal calculation extracts $\IR^\H_{am}$ as a curvature component of the total-space connection and then uses the $[1,3]$ component of \eqref{eq:universal-system}, whereas the direct calculation of \citeAPmetric varies the composite field $\Th^\H(g,H)$, including the induced vielbein variation and the gauge and Lorentz Chern--Simons terms.  Agreement of the tensor structure and relative coefficients checks both the universal deformation calculus and the identification of $\ITheta^\H$ as the universal Hull connection.

The analysis of \citeAP shows that the curvature  $\ap R^{\H\,0,2}\= \cO(\ap^2)$ and so, unlike the gauge connection, we can no longer treat the tangent bundle connection in the framework of a holomorphic structure at $\ap^2$.  Since $\IR^\H$ restricts fibrewise to $R^\H$, the vertical $(0,2)$ component of the universal curvature is then nonzero.  One might worry that this also forces mixed anti-holomorphic components such as $\fD_{\bar\a}\Th^\H_{\bar\m}$ to be nonzero.  Subsection~\ref{s:IRH_Hol} shows that it does not.  For the Hull connection,
\beq\label{eq:Hullhol}
\ap \IR^\H_{\ab\mb}\=\ap \fD_{\ab}\Th^\H_{\mb}\= 0+\cO(\ap^3)~.
\eeq
This is holomorphy in the moduli directions; it does not imply
$(R^\H)^{0,2}=0$ on the fibre. Thus the Hull connection continues to depend holomorphically on parameters.  This is a nontrivial check that the universal geometry remains compatible with $\ap$ corrections and with the composite nature of $R^\H$.

This should be contrasted with the gauge connection $\IA$.  For the gauge bundle, the horizontal component $A_a^{\sharp}$ is part of the choice of lift from $\M$ to the space of fields.  Once $F^{0,2}=0$ holds on each fibre, the mixed component of $\IF^{0,2}=0$ can therefore be read as a holomorphic gauge choice for the deformation representatives. The Hull connection has a different status.  It is composite, so its variation along $\M$ is fixed by the universal metric and $B$-field deformations.  Hence \eqref{eq:Hullhol} is not an additional gauge condition on $\ITheta^\H$, but the result of the computation. 

Only the F-terms and Green--Schwarz identity, provided Hull curvature is used, extend to the total space. We impose the D-terms $\L_\o F = 0$ and $\dd (\norm{\O}\o^2 ) = 0$ fiberwise. Extending them to $\IX$ would require extra data, such as a metric or contraction operator on the total space. It would also overly constrain the mixed components $\IF_{am}$ and $\IR^\H_{am}$, which describe deformations and need not vanish. Thus there is no natural tangent-bundle instanton equation on $\IX$. The object that does lift is the composite Hull connection $\ITheta^\H(g,H)$, since $g$ and $H$ already have universal extensions.

The same comment applies to the holomorphic three-form.  The form $\O$ is a section of the relative canonical bundle $K_{\IX/\M}$, not a holomorphic volume form on $\IX$; a volume form on the total space would have degree $3+\dim_{\mathbb C}\M$ and would require extra base data.  The fibre form is closed only vertically: schematically,
\beq\label{eq:relative-Omega}
\dd_X\O_y\=0~,\qquad
\Id\O\=\dd y^\a\w\fD_\a\O+\dd y^{\ab}\w\fD_{\ab}\O~.
\eeq
For a nontrivial family the mixed terms encode the variation of Hodge structure and need not vanish.

A  concrete check of universal geometry comes from the heterotic moduli-space metric $g_{\a\bb}^\sharp$ and its \K potential $K$ \cite{Candelas:2016usb,McOrist:2019mxh,McOrist:2026cys}:
\beq\label{eq:Kpotential-letter}
g^{\sharp}_{\a\bb}\=\del_\a\del_\bb K+\cO(\ap^3)~,
\qquad
K\=\!-\log\left(\frac43\int_X\wh \o^3\right)  -\log\left(\ii\int_X\O\w\Ob\right) + \cO(\ap^3) ~.
\eeq
Despite the simple form of $K$, its ability to capture the gauge-bundle dependence in the moduli-space metric rests on a set of complicated second-order algebraic relations, all encoded by the universal Bianchi identity \cite{Candelas:2016usb,Candelas:2018lib}.  At order $\ap$, the metric passes another nontrivial test: it defines an adjoint $\Dbar^\dag$ on the extension bundle $Q$ whose kernel gives the D-term representative equations \cite{McOrist:2021dnd,McOrist:2024zdz}.  At order $\ap^2$, the same \K potential reproduces the moduli-space metric obtained by dimensional reduction of the $\ap^2$-corrected Bergshoeff--de Roo action \cite{Bergshoeff:1989de}.

\section{The moduli space of the moduli space problem}
This agreement is already a strong result. The unchanged form of the universal F-term and anomaly equations has a further, surprising payoff: their infinitesimal deformations are governed by the same extension structure and surprisingly give direct access to second-order deformation theory \cite{McOrist:2024glz}.


A universal tensor with one moduli leg already represents a first covariant derivative of a fibre field. Deforming the universal geometry therefore gives second covariant derivatives of the original fibre data. For example, varying $\IF^{0,2}=0$ constrains $\fD_a\fD_bA$, while varying $N_\IJ=0$ constrains $\fD_a\fD_bJ$.  A direct calculation must track changes of type, connections and gauge representatives. The universal equations collect these terms into a small set of tensor equations.

So far, this calculation has been carried out only through order $\ap$. It uses an auxiliary ${\rm End}\,\ccT_\IX$ summand to represent the induced variation of the universal Hull connection; this summand is not an independent modulus. The $\ap^2$ calculation requires a reduced universal operator, which has not yet been constructed.

To carry out the order-$\ap$ calculation, introduce the auxiliary universal bundle $\IQ_{\rm aux}$ on $\IX$, with fibre
\beq\label{eq:Quniv}
(\ccT_\IX^{1,0})^*\oplus {\rm End}\,\ccT_\IX
\oplus {\rm End}\,\ccU \oplus\ccT_\IX^{1,0}~.
\eeq
 We use the following upper triangular matrix to organise the order-$\ap$ calculation:
\beq\label{eq:Dbar-univ}
\IDb\=
\begin{pmatrix}
\Idelb &- \ap \IR^* & \ap \IF^* & \IH\\
0 & \Idelb_{\Th^\H} & 0 &  \wt\IR\\
0 & 0 & \Idelb_\IA &  \wt\IF\\
0 & 0 & 0 & \Idelb
\end{pmatrix}~,
\eeq
Here $\wt \IF(\d \IJ) = \d \IJ^M \IF_M$ and $\wt \IR(\d \IJ) = \d \IJ^M \IR_M$, while $\IF^*(\d \IA^{0,1}) = \half \tr (\IF \w \d \IA^{0,1})$, $\IR^*(\d \ITheta^{0,1}) = \half \tr (\IR \w \d \ITheta^{0,1})$ and $\IH(\d \IJ) = -\ii \IH^{1,1}_M \d \IJ^M$.  Thus the off-diagonal entries in \eqref{eq:Dbar-univ} are the universal Atiyah maps and their duals, together with the flux map. We remark that with the normalisation inherited from the physical theory, $\ap \Idelb_{\ITheta^\H}^{\,2} = \ap \IR^{\H\,0,2} = \cO(\ap^2)$.  This is a reason \eqref{eq:Dbar-univ} is used  only to organise the order-$\ap$ calculation; at next order we need to dispense with the  tangent bundle component.   

The universal deformation equations then constrain the components with at least two-moduli space legs.  With some  assumptions---a smooth patch of $\M$, an integrable horizontal distribution, holomorphic gauge, absence of infinitesimal bundle automorphisms $H^0(X,{\rm End}\,E)=0$ and preservation of the D-terms on every fibre---one obtains, through order $\ap$,
\beq\label{eq:flatness-letter}
\IF_{ab}\=0~,
\qquad
\IH_{ab}\=0~,
\qquad
\IH_{abc}\=0~,
\eeq
for the pure moduli-space components relevant here.  Equation \eqref{eq:flatness-letter} does not set the physical deformation $\IF_{am}=\fD_aA_m$ to zero.  It states that the connection used to compare those deformations is locally flat once the fibrewise Hermitian Yang--Mills equation and stability are imposed.  Many second derivatives of $J$ and $A$ consequently vanish in holomorphic type, while the remaining mixed derivatives are fixed by products of first-order deformations and the Atiyah maps.  At this order the D-terms therefore enter as the mechanism behind flat transport over the moduli space.

As an example of how this works, the universal F-term equations and holomorphic gauge imply, among other relations, the identities
\beq\label{eq:second-derivative-letter}
\fD_\bb\fD_\a\A\=0~,
\qquad
\fD_\a\fD_\bb\A\=\delb_\A\IF_{\a\bb}~,
\qquad
\fD_\a\fD_\bb J^\m\=0~.
\eeq
These identities can be verified by direct calculation.  At this stage the pure moduli-space curvature $\IF_{\a\bb}$ need not vanish.  Requiring the universally deformed connection to continue to satisfy the fibrewise Hermitian Yang--Mills equation gives $\Box_{\delb_\A}\IF_{\a\bb}=0$.  Stability of $E$ then implies $\IF_{\a\bb}=0$, up to possible central components and hence $\fD_\a\fD_\bb\A=0$.

The same universal calculation leaves nontrivial $(1,1)$ components of the curvature $\IR_{\a\bb}$: these include the curvature of the \K metric on $\M$ and terms built from the fibre complex-structure and Hermitian deformations.  The contrast with the gauge sector is expected.  In the gauge sector, stability and the fibrewise D-terms force the pure moduli-space curvature $\IF_{\a\bb}$ to vanish locally.  In the tangent-bundle sector, however, the connection is composite, so its curvature also records the variation of the underlying metric and three-form.  This is reminiscent of the gauge--Lorentz correspondence of \cite{Bergshoeff:1989de}: at leading order the torsionful tangent-bundle connection can be organised like a Yang--Mills multiplet, but beyond leading order the correspondence is corrected because the Lorentz connection is induced from supergravity fields rather than chosen independently.  This is precisely why the auxiliary calculation cannot be extrapolated unchanged to order $\ap^2$.

\section{Toward heterotic special geometry}
 Even at order $\ap$, universal geometry supplies both ingredients needed for a heterotic special geometry: mixed curvatures determine infinitesimal deformations and their $L^2$ metric, while curvatures along $\M$ determine how those deformations are compared at neighbouring points. Any heterotic analogue of special geometry should contain not only tangent spaces and their metric, but also a rule for parallel transport.  At order $\ap$, the classes in $H^1_{\Dbar}(X,Q)$ describe infinitesimal F-term deformations at a fixed compactification, with the D-terms selecting physical representatives.  Physical couplings require comparing those classes and representatives as $y$ varies.  The universal connection provides this comparison map: its  curvature along the moduli space controls commutators of covariant derivatives and therefore the structure of higher-point couplings.  In Calabi--Yau special geometry the corresponding roles are supplied by the Gauss--Manin connection and variation of Hodge structure.  The Atiyah and anomaly maps mix the sectors, so heterotic compactifications do not admit the same Hodge-theoretic factorisation as Calabi--Yau moduli spaces. Universal connections nevertheless supply the data needed to define parallel transport.

The $\ap^2$ corrections test this picture: the universal construction remains compatible with the corrected F-term and anomaly equations of \citeAP.  The  equations retain their leading form and lift naturally to the total space with the composite connection $\ITheta^\H(g,H)$.  The D-terms are imposed fibrewise and the leading-order instanton property of $R^\H$ can fail perturbatively.  Thus the natural connection appearing in the universal Green--Schwarz identity is the Hull connection itself, viewed as a composite of the corrected fields.  

Through order $\ap$, the Green--Schwarz Bianchi identity supplies the integrability condition for the deformation complex, $\Dbar^2=\cO(\ap^2)$.  Universal geometry packages this identity and its variations over the entire family, while small-gauge fixing is encoded by the choice of holomorphic structures and horizontal distributions. The auxiliary universal bundle $\IQ_{\rm aux}$ suggests that a universal analogue of the reduced bundle $Q$ should govern higher jets of the heterotic family. Constructing that reduced complex through order $\ap^2$, and determining whether the resulting hierarchy closes into a flat, $L_\infty$, or variation-of-Hodge-type structure, remain open questions.

Two limitations are important. The F-term and anomaly equations lift through $\ap^2$ but the  double-extension calculation and the associated flatness statements are known at $\ap$.  Extending them to order $\ap^2$ requires eliminating the auxiliary ${\rm End}\,\ccT_\IX$ directions, constructing the reduced universal bundle and its corrected differential, and finding the corresponding adjoint.  Global patching of the universal geometry and its behaviour near singular loci are separate open problems.  Resolving these issues is necessary before the local flatness structure can be promoted to a global heterotic special geometry.

Universal geometry may also constrain higher-order $\ap$ corrections. The same universal equations control the \K property of the moduli-space metric, holomorphic gauge, second-order deformation constraints and flat transport in the gauge-bundle directions. This may help identify which higher-order corrections preserve heterotic supersymmetry and anomaly cancellation.

\vskip0.5cm
\paragraph*{Acknowledgements.}
I would like to thank J. Knapp, S. Picard and E. Svanes for enlightening conversations and very helpful comments.  JM is  supported in part by ARC Discovery Project Grants DP240101409 and DP250101828.

\appendix

\section{The Hull curvature and the first \texorpdfstring{$\ap$}{alpha-prime} correction}
\label{app:mixed-Hull-curvature}

This appendix reproduces the mixed-tangibility calculation used in Section~4 of \cite{Candelas:2018lib} and then  computes its first order $\ap$-correction.  We work in a large-radius expansion and compute the universal Hull curvature to first order in $\ap$. This is sufficient to determine the $\ap^2$ deformation equations. 

We assume $S_{ab}{}^m=0$.  Locally, in real coordinates on the fibre, this allows a product gauge
\beq
 c_a{}^m=0,
 \qquad
 e_a=\del_a,
 \qquad
 e^m=\dd x^m .
 \label{eq:app-product-gauge}
\eeq
This real gauge is only a convenient way to see that the calculation is tensorial.  It should not be confused with choosing holomorphic coordinates on a family with varying complex structure.  In holomorphic fibre coordinates one must keep the derivatives of the horizontal shift, since
\beq
 \D_{\alpha\mb}{}^\nu=-\del_{\mb}c_\alpha{}^\nu
 \label{eq:app-Delta-c}
\eeq
is the Kodaira--Spencer tensor for first order deformations in complex structure.

Write the universal Hull connection in the vertical tangent directions as
\beq
 \ITheta^\H
 =\dd x^m\,\Th^\H_m+\dd y^a\,\ITheta^\H_a .
 \label{eq:app-universal-connection-real}
\eeq
Its curvature is
\beq
 \IR^\H=\Id\ITheta^\H+\ITheta^\H\w\ITheta^\H .
 \label{eq:app-universal-curvature-def}
\eeq
Taking the coefficient of $\dd y^a\w\dd x^m$ gives
\beq
 (\IR^\H)_{am}
 =\del_a\Th^\H_m-\nabla^\H_m\ITheta^\H_a .
 \label{eq:app-mixed-curvature-real}
\eeq
Equivalently,
\beq
 \fD_a\Th^\H=(\IR^\H)_{am}\,\dd x^m .
 \label{eq:app-DTheta-from-R}
\eeq
Thus $\fD_a\Th^\H$ is a mixed component of the corresponding universal curvature tensor.

\subsection{The holomorphic mixed component}
\label{s:IRH_Hol}

At zeroth order in $\ap$ the computation is the one in Section~4 of \cite{Candelas:2018lib}.  After carrying out the real product-gauge calculation, choose holomorphic indices on a fixed fibre and set
\beq
 K_\alpha{}^\nu{}_{\sigma}
 = g^{\nu\bar\lambda}\fD_\alpha g_{\sigma\bar\lambda} .
 \label{eq:app-K-def}
\eeq
Using $\IH=\Id^c\wh\Iomega$, the component of the connection given in
Section~4.2 of \citeUG is
\beq
 \ITheta^\H_\alpha{}^\nu{}_{\sigma}\=K_\alpha{}^\nu{}_{\sigma} + \cO(\ap^2).
 \label{eq:app-hol-horizontal}
\eeq
Moreover, for the holomorphic fibre component one has
\beq
 \Th^\H_\mu{}^\nu{}_{\sigma}
 =(g^{-1}\del_\mu g)^\nu{}_{\sigma} .
 \label{eq:app-hol-fibre-Hull}
\eeq
Hence
\beq
 \del_\alpha(g^{-1}\del_\mu g)
 =\del_\mu(g^{-1}\del_\alpha g)
 +[g^{-1}\del_\mu g,\,g^{-1}\del_\alpha g] ,
 \label{eq:app-matrix-maurer-cartan}
\eeq
or, in covariant notation,
\beq
 \del_\alpha\Th^\H_\mu{}^\nu{}_{\sigma}\=\nabla_\mu K_\alpha{}^\nu{}_{\sigma} +\cO(\ap^2)~.
 \label{eq:app-matrix-identity}
\eeq
Substitution into the mixed curvature gives
\beq
\begin{split}
 (\IR^\H_\alpha)_\mu{}^\nu{}_{\sigma} &\=\del_\alpha\Th^\H_\mu{}^\nu{}_{\sigma}  -\nabla_\mu K_\alpha{}^\nu{}_{\sigma}+\cO(\ap^2) \\[5pt]
 &\=\cO(\ap^2)~.
\end{split}
 \label{eq:app-p34-result}
\eeq
This reproduces the final equation on p.~34 of \cite{Candelas:2018lib} for the Hull connection.  Since the supersymmetry relation $\IH=\Id^c\wh\Iomega$ is unchanged in form, the same algebraic cancellation holds after the first $\ap$ corrections to the background fields are inserted.  Thus
\beq
 (\IR^\H_\alpha)_\mu{}^\nu{}_{\sigma} \=\fD_\alpha\Th^\H_\mu{}^\nu{}_{\sigma}\= \cO(\ap^2)~,
 \label{eq:app-hol-mixed-zero}
\eeq
for the Hull connection. As the terms discussed here come with an explicit $\ap$, we only need to work to $\ap^2$ as the remainder contributes at $\ap^3$. From now on we suppress the $\cO(\ap^2)$ except where it is needed for clarity. 

This should be distinguished from the fibre instanton condition.  The corrected fibre curvature may have $(R^\H)^{0,2}\ne0$.  This gives a nonzero purely vertical component of $\IR^\H$, because $\IR^\H|_{X_y}=R^\H$, but it does not force the mixed holomorphic component \eqref{eq:app-hol-mixed-zero} to become nonzero.  The universal bundle therefore accommodates the corrected Hull curvature without producing spurious mixed curvature components.

\subsection{The mixed component \texorpdfstring{$\IR^\H_{\alpha\mb}$}{IRH alpha mbar}}
\label{s:mixedsub}
We next compute
\beq
 \IR^\H_{\alpha\mb}=\fD_\alpha\Th^\H_{\mb} .
\eeq
This component gives the deformation of the $(0,1)$ part of the Hull connection. The calculation must keep the complex-structure variation: although $c_a{}^m$ can be set to zero in the real product gauge, its holomorphic derivatives reappear through \eqref{eq:app-Delta-c}.

Define
\beq
 K_{\alpha\rho}{}^\nu
 =g^{\nu\bar\lambda}\fD_\alpha g_{\rho\bar\lambda},
 \qquad
 \ccZ_\alpha^{(1,1)}=2\ii\,\fD_\alpha\omega^{(1,1)} .
 \label{eq:app-K-Z-def}
\eeq
Since $\omega_{\rho\bar\lambda}=\ii g_{\rho\bar\lambda}$, this implies
\beq
 K_\a{}^\n{}_\r \=\!-\half g^{\nu\bar\lambda}\ccZ_{\alpha\rho\bar\lambda} .
 \label{eq:app-K-Z-relation}
\eeq
For the Hull connection the relevant component on the total space is
\beq
 \ITheta^\H_\alpha{}^\nu{}_{\rho} \=\del_\rho c_\alpha{}^\nu+ K_\a{}^\n{}_\r ~,
 \label{eq:app-horizontal-Hull-component}
\eeq
while the fibre component obeys
\beq
  \Th^\H_{\mb}{}^\nu{}_{\rho} \= 0~.
 \label{eq:app-Hull-zero-component}
\eeq

Taking the $\dd y^\alpha\w e^{\mb}$ coefficient of
$\IR^\H=\Id\ITheta^\H+(\ITheta^\H)^2$ gives
\beq
\begin{split}
 (\IR^\H_{\alpha\mb})^\nu{}_{\rho}
 &\= e_\alpha(\Th^\H_{\mb}{}^\nu{}_{\rho})
    -(\del_{\mb}c_\alpha{}^\sigma)\Th^\H_\sigma{}^\nu{}_{\rho}
    -\nabla^\H_{\mb}(\ITheta^\H_\alpha{}^\nu{}_{\rho}) \\
 &\= \D_{\alpha\mb}{}^\sigma\Th^\H_\sigma{}^\nu{}_{\rho}
    -\nabla^\H_{\mb}\del_\rho c_\alpha{}^\nu
    -\nabla^\H_{\mb}K_\a{}^\n{}_\r ~.
\end{split}
 \label{eq:app-full-mixed-Hull}
\eeq
In the first line the term $e_\alpha(\Th^\H_{\mb}{}^\nu{}_{\rho})$ vanishes by \eqref{eq:app-Hull-zero-component}; the second term is the moving-frame contribution.  We now simplify the two pieces in \eqref{eq:app-full-mixed-Hull}.  First, convert the Hull covariant derivative of $\del_\rho c_\alpha{}^\nu$ to Levi--Civita:
\beq
\begin{split}
 -\nabla^\H_{\mb}\del_\rho c_\alpha{}^\nu
 &\=\!-\nabla_{\mb}\del_\rho c_\alpha{}^\nu
   -\half H_{\mb}{}^\nu{}_{\sigma}\del_\rho c_\alpha{}^\sigma
   +\half H_{\mb}{}^\sigma{}_{\rho}\del_\sigma c_\alpha{}^\nu~.
\end{split}
 \label{eq:app-H-to-LC-shift}
\eeq
Using the Levi--Civita symbols, for example as in Appendix~C of \citeAP, one has
\beq
\begin{split}
 -\nabla_{\mb}\del_\rho c_\alpha{}^\nu
 &\=\!-\del_\rho\del_{\mb}c_\alpha{}^\nu
   +\half H_{\mb}{}^\nu{}_{\sigma}\del_\rho c_\alpha{}^\sigma
   -\half H_{\mb}{}^\sigma{}_{\rho}\del_\sigma c_\alpha{}^\nu~.
\end{split}
 \label{eq:app-LC-shift}
\eeq
The $H$-terms in \eqref{eq:app-H-to-LC-shift} and \eqref{eq:app-LC-shift}
cancel and therefore
\beq
 -\nabla^\H_{\mb}\del_\rho c_\alpha{}^\nu
 \=\!-\del_\rho\del_{\mb}c_\alpha{}^\nu
 \=\del_\rho\D_{\alpha\mb}{}^\nu .
 \label{eq:app-shift-cancel}
\eeq
The first two terms of \eqref{eq:app-full-mixed-Hull} become
\beq
  e_\alpha(\Th^\H_{\mb}{}^\nu{}_{\rho})
    -(\del_{\mb}c_\alpha{}^\sigma)\Th^\H_\sigma{}^\nu{}_{\rho}\= \del_\rho\D_{\alpha\mb}{}^\nu
 +\D_{\alpha\mb}{}^\sigma\Th^\H_\sigma{}^\nu{}_{\rho}~.
 \label{eq:app-c-part-raw}
\eeq
Using $\Th^\H_\sigma{}^\nu{}_{\rho}=\Th^\Ch_\sigma{}^\nu{}_{\rho}=\Th^\LC_\s{}^\n{}_\r - \half H_\s{}^\n{}_\r$, we can write the right hand side in a manifestly covariant manner
\beq
\begin{split}
 \del_\rho\D_{\alpha\mb}{}^\nu
 +\D_{\alpha\mb}{}^\sigma\Th^\H_\sigma{}^\nu{}_{\rho}
 &\=\nabla_\rho\D_{\alpha\mb}{}^\nu
   +\ii g^{\nu\bar\lambda}
     \big(\D_\alpha{}^\sigma(\del\omega)_\sigma\big)_{\rho\mb\bar\lambda} + \cO(\ap^2)~,
\end{split}
 \label{eq:app-c-part-tensorial}
\eeq
where we use $H=\ii(\del-\delb)\o$, $\D_{\a[\mb\nb]}=\cO(\ap)$ and $H=\cO(\ap)$.

Finally, using \eqref{eq:app-K-Z-relation} and metric compatibility of the Hull
connection,
\beq
 -\nabla^\H_{\mb}K_\a{}^\n{}_\r \=\half g^{\nu\bar\lambda}\nabla_{\mb}\ccZ_{\alpha\rho\bar\lambda} .
 \label{eq:app-K-term}
\eeq
Combining \eqref{eq:app-full-mixed-Hull}, \eqref{eq:app-c-part-tensorial} and
\eqref{eq:app-K-term} gives
\beq
\begin{split}
 (\IR^\H_{\alpha\mb})^\nu{}_{\rho}
 &=\nabla_\rho\D_{\alpha\mb}{}^\nu +\half\nabla^\nu\ccZ_{\alpha\rho\mb} -\half g^{\nu\bar\lambda} \left(\delb\ccZ_\alpha -2 \ii \D_\alpha{}^\sigma(\del\omega)_\sigma\right)_{\rho\mb\bar\lambda} + \cO(\ap^2)~.
\end{split}
 \label{eq:app-mixed-R-before-bianchi}
\eeq
where we used
\beq
 (\delb\ccZ_\alpha)_{\rho\mb\bar\lambda}
 =\nabla_{\bar\lambda}\ccZ_{\alpha\rho\mb}
  -\nabla_{\mb}\ccZ_{\alpha\rho\bar\lambda} .
 \label{eq:app-delbZ-convention}
\eeq
in \eqref{eq:app-K-term}.  The $[1,3]$ component of the universal Green--Schwarz identity gives
\beq
\begin{split}
 (\delb\ccZ_\alpha)_{\rho\mb\bar\lambda}
 &=2\ii\,\big(\D_\alpha{}^\sigma(\del\omega)_\sigma\big)_{\rho\mb\bar\lambda}    \\
 &\quad
 +\frac{\ap}{2}
 \left[\tr(\aa_\alpha\w F)-\tr(\fD_\alpha\Th^\H\w R^\H)\right]_{\rho\mb\bar\lambda} .
\end{split}
 \label{eq:app-Z-bianchi-component}
\eeq
Substituting \eqref{eq:app-Z-bianchi-component} into \eqref{eq:app-mixed-R-before-bianchi} leaves
\beq
\begin{split}
 (\IR^{\H}_{\alpha\mb}){}^\nu{}_{\rho}\= (\fD_\a \Th^{\H\,0,1})^\n{}_\r
 &=\nabla_\rho\D_{\alpha\mb}{}^\nu
 +\half\nabla^\nu\ccZ_{\alpha\rho\mb}  \\
 &\quad
 -\frac{\ap}{4}\,g^{\nu\bar\lambda}
 \left[\tr(\aa_\alpha\w F)-\tr(\fD_\alpha\Th^\H\w R^\H)
 \right]_{\rho\mb\bar\lambda} + \cO(\ap^2)~.
\end{split}
 \label{eq:app-leading-47}
\eeq
At $\ap=0$ this reduces to equation~(4.7) of \cite{Candelas:2018lib}.  Solving perturbatively in $\ap$ gives
\beq
\begin{split}
 \fD_\a\Theta^{\H\,0,1}{}^\n{}_{\m} & \= \nabla_\m \D_\a^\n +\half \nabla^\n \ccZ_{\a\,\m}
-\frac{\ap}{12}\tr\left(\aa \,F_{\m\lb}+\aa_\lb F_\m\right)g^{\lb\n}\\[3pt]
&
\quad+\frac{\ap}{6}\Bigg( \left(\nabla_\s \D^\t + \half \nabla^\t \ccZ_{\s}\right)R^\H{}_{\m}{}^{\n\s}{}_\t + \left(\nabla_\s \D_{\tb}{}^\n +\half \nabla_\tb \ccZ_{ \s}{}^\n \right)R^\H{}_{\m}{}^{\s\tb}\Bigg) + \cO(\ap^2)~,\\[3pt]
\end{split}
\eeq
This agrees exactly with the deformation calculation of \citeAPmetric. The universal curvature reproduces the tensor structure obtained by varying the composite field $\Th^\H(g,H)$ directly. In particular, the tangent-bundle connection remains tied to the Hull connection rather than being treated as an independent instanton connection.

\subsection{The off-diagonal tangent directions}
\label{s:app-off-diagonal-Hull-blocks}

It remains to record the two vertical tangent directions
$(\IR^\H_{\alpha\lb})^{\nb}{}_{\rho}$ and
$(\IR^\H_{\alpha\lb})^{\nu}{}_{\rb}$.  We use the same universal-curvature
calculation as in \eqref{eq:app-full-mixed-Hull}, now with different matrix
indices.  

The relevant coefficient of
$\IR^\H=\Id\ITheta^\H+(\ITheta^\H)^2$ is
\beq
 (\IR^\H_{\alpha\lb})_{mn}
 =(\Dethsharp_\alpha \Th^\H_{\lb})_{mn}
  -(\nabla^\H_{\lb}\ITheta^\H_\alpha)_{mn}~.
 \label{eq:app-offdiag-curvature-coeff}
\eeq
Here $\Dethsharp_\alpha$ includes the moving-frame term from the
$e^{\lb}$ one-form index, exactly as in \eqref{eq:app-full-mixed-Hull}.

Set $(\e,\r)=(1,0)$ in the symbol list of Section~4.2 of \citeUG.  The  symbols needed below are
\beq
\begin{split}
 (\ITheta^\H_\alpha)_{\m\rho}&\=0~,
        \qquad
 (\Th^\H_{\lb})_{\m\rho}=H_{\lb\m\rho}~, \\[3pt]
 (\ITheta^\H_\alpha)_{\mb\rb}&\=2\D_{\alpha[\mb\rb]}~,
        \qquad
 (\Th^\H_{\lb})_{\mb\rb}\=0~.
\end{split}
 \label{eq:app-offdiag-symbols}
\eeq
For $m,n$ both holomorphic, the first component, after using
\eqref{eq:app-offdiag-symbols}, gives
\beq
 (\IR^\H_{\alpha\lb})_{\m\rho}
   =\Dethsharp_\alpha H_{\lb\m\rho}~.
 \label{eq:app-offdiag-top-as-H}
\eeq
The identity $\IH=\Id^c\wh\Iomega$ gives, in this component, $H_{\lb\m\rho}=\del_\rho g_{\m\lb}-\del_\m g_{\rho\lb} + \cO(\ap^2)$.  Applying the horizontal derivative, the non-tensorial pieces combine into covariant derivatives of the metric deformation:
\beq
\begin{split}
 \Dethsharp_\alpha H_{\lb\m\rho}
 &\=\nabla_\rho(\fD_\alpha g_{\m\lb})
   -\nabla_\m(\fD_\alpha g_{\rho\lb})   \=\half\left(
      \nabla_\m\ccZ_{\alpha\rho\lb}
     -\nabla_\rho\ccZ_{\alpha\m\lb}
    \right)~.
\end{split}
 \label{eq:app-offdiag-Z-block-lowered}
\eeq
For the case where both $m,n$ are both anti-holomorphic,  $(\Th^\H_{\lb})_{\mb\rb}=0$, and hence
\eqref{eq:app-offdiag-curvature-coeff} gives
\beq
\begin{split}
 (\IR^\H_{\alpha\lb})_{\mb\rb}
 &=-\nabla^\H_{\lb}(\ITheta^\H_\alpha)_{\mb\rb}
   =-2\nabla_{\lb}\D_{\alpha[\mb\rb]}~.
\end{split}
 \label{eq:app-offdiag-bottom-first}
\eeq
The Hull and Levi--Civita derivatives agree here because there is no $(0,3)$ component of $H$.  Using $\delb\D_\alpha=0$, so that $\nabla_{\lb}\D_{\alpha\mb\rb} =\nabla_{\mb}\D_{\alpha\lb\rb}$ we find
\beq
\begin{split}
 (\IR^\H_{\alpha\lb})_{\mb\rb}
   \=\!-\nabla_{\mb}\D_{\alpha\lb\rb}
   +\nabla_{\rb}\D_{\alpha\lb\mb}~.
\end{split}
 \label{eq:app-offdiag-Delta-block-lowered}
\eeq
Putting it together,  writing $\ccZ_{\alpha\m}=\ccZ_{\alpha\m\lb}e^{\lb}$, $\D_{\alpha\mb}=\D_{\alpha\lb\mb}e^{\lb}$, we find
\beq
\begin{split}
 \fD_\alpha\Th^{\H\,0,1}_{\m\rho}
 \=\half\left(
      \nabla_\m\ccZ_{\alpha\rho}
     -\nabla_\rho\ccZ_{\alpha\m}
    \right), \qquad  \fD_\alpha\Th^{\H\,0,1}_{\mb\rb}
\=\!-\nabla_\mb\D_{\alpha\rb}
   +\nabla_\rb\D_{\alpha\mb}~.
\end{split}
 \label{eq:app-offdiag-metric-paper-form}
\eeq
These are the first and last tangent-index directions in the deformation formula
used in \citeAPmetric.  Since $\del\ccZ_\alpha=\cO(\ap)$ and
$\D_{\alpha[\mb\rb]}=\cO(\ap)$, both terms vanish in the Kähler
$\ap\to0$ limit, as expected.


\begin{thebibliography}{10}

\bibitem{Hull:1986kz}
C.~Hull, \emph{{Compactifications of the Heterotic Superstring}},
  \href{https://doi.org/10.1016/0370-2693(86)91393-6}{\emph{Phys.Lett.}
  {\bfseries B178} (1986) 357}.

\bibitem{Bergshoeff:1989de}
E.~Bergshoeff and M.~de~Roo, \emph{{The Quartic Effective Action of the
  Heterotic String and Supersymmetry}},
  \href{https://doi.org/10.1016/0550-3213(89)90336-2}{\emph{Nucl.Phys.}
  {\bfseries B328} (1989) 439}.

\bibitem{McOrist:2025zwf}
J.~McOrist and S.~Picard, \emph{{Stringy Corrections to Heterotic
  SU(3)-Geometry}},
  \href{https://doi.org/10.1007/s00220-026-05620-6}{\emph{Commun. Math. Phys.}
  {\bfseries 407} (2026) 105}
  [\href{https://arxiv.org/abs/2507.02388}{{\ttfamily 2507.02388}}].

\bibitem{Strominger:1986uh}
A.~Strominger, \emph{{Superstrings with Torsion}},
  \href{https://doi.org/10.1016/0550-3213(86)90286-5}{\emph{Nucl. Phys. B}
  {\bfseries 274} (1986) 253}.

\bibitem{delaOssa:2014cia}
X.~de~la Ossa and E.~E. Svanes, \emph{{Holomorphic Bundles and the Moduli Space
  of N=1 Supersymmetric Heterotic Compactifications}},
  \href{https://doi.org/10.1007/JHEP10(2014)123}{\emph{JHEP} {\bfseries 10}
  (2014) 123} [\href{https://arxiv.org/abs/1402.1725}{{\ttfamily 1402.1725}}].

\bibitem{Anderson:2014xha}
L.~B. Anderson, J.~Gray and E.~Sharpe, \emph{{Algebroids, Heterotic Moduli
  Spaces and the Strominger System}},
  \href{https://doi.org/10.1007/JHEP07(2014)037}{\emph{JHEP} {\bfseries 07}
  (2014) 037} [\href{https://arxiv.org/abs/1402.1532}{{\ttfamily 1402.1532}}].

\bibitem{Candelas:2016usb}
P.~Candelas, X.~de~la Ossa and J.~McOrist, \emph{{A Metric for Heterotic
  Moduli}}, \href{https://doi.org/10.1007/s00220-017-2978-7}{\emph{Commun.
  Math. Phys.} {\bfseries 356} (2017) 567}
  [\href{https://arxiv.org/abs/1605.05256}{{\ttfamily 1605.05256}}].

\bibitem{McOrist:2021dnd}
J.~McOrist and E.~E. Svanes, \emph{{Heterotic quantum cohomology}},
  \href{https://doi.org/10.1007/JHEP11(2022)096}{\emph{JHEP} {\bfseries 11}
  (2022) 096} [\href{https://arxiv.org/abs/2110.06549}{{\ttfamily
  2110.06549}}].

\bibitem{McOrist:2024zdz}
J.~McOrist, S.~Picard and E.~E. Svanes, \emph{{A Heterotic Hermitian-Yang-Mills
  Equivalence}},
  \href{https://doi.org/10.1007/s00220-025-05272-y}{\emph{Commun. Math. Phys.}
  {\bfseries 406} (2025) 107}
  [\href{https://arxiv.org/abs/2402.10354}{{\ttfamily 2402.10354}}].

\bibitem{Hull:1986xn}
C.~M. Hull and P.~K. Townsend, \emph{{World Sheet Supersymmetry and Anomaly
  Cancellation in the Heterotic String}},
  \href{https://doi.org/10.1016/0370-2693(86)91493-0}{\emph{Phys. Lett. B}
  {\bfseries 178} (1986) 187}.

\bibitem{Ross:1986ra}
D.~A. Ross, \emph{{Chern-simons Terms in the $\sigma$ Model for the Heterotic
  String}}, \href{https://doi.org/10.1016/0550-3213(87)90433-0}{\emph{Nucl.
  Phys. B} {\bfseries 286} (1987) 93}.

\bibitem{Hull:1987pc}
C.~M. Hull and P.~K. Townsend, \emph{{The Two Loop Beta Function for $\sigma$
  Models With Torsion}},
  \href{https://doi.org/10.1016/0370-2693(87)91331-1}{\emph{Phys. Lett. B}
  {\bfseries 191} (1987) 115}.

\bibitem{Metsaev:1986yb}
R.~R. Metsaev and A.~A. Tseytlin, \emph{{Curvature Cubed Terms in String Theory
  Effective Actions}},
  \href{https://doi.org/10.1016/0370-2693(87)91527-9}{\emph{Phys. Lett. B}
  {\bfseries 185} (1987) 52}.

\bibitem{Cai:1986sa}
Y.~Cai and C.~A. Nunez, \emph{{Heterotic String Covariant Amplitudes and
  Low-energy Effective Action}},
  \href{https://doi.org/10.1016/0550-3213(87)90106-4}{\emph{Nucl. Phys. B}
  {\bfseries 287} (1987) 279}.

\bibitem{Metsaev:1987zx}
R.~Metsaev and A.~A. Tseytlin, \emph{{Order alpha-prime (Two Loop) Equivalence
  of the String Equations of Motion and the Sigma Model Weyl Invariance
  Conditions: Dependence on the Dilaton and the Antisymmetric Tensor}},
  \href{https://doi.org/10.1016/0550-3213(87)90077-0}{\emph{Nucl. Phys. B}
  {\bfseries 293} (1987) 385}.

\bibitem{Gross:1986mw}
D.~J. Gross and J.~H. Sloan, \emph{{The Quartic Effective Action for the
  Heterotic String}},
  \href{https://doi.org/10.1016/0550-3213(87)90465-2}{\emph{Nucl. Phys. B}
  {\bfseries 291} (1987) 41}.

\bibitem{Foakes:1988wy}
A.~P. Foakes, N.~Mohammedi and D.~A. Ross, \emph{{Three Loop Beta Functions for
  the Superstring and Heterotic String}},
  \href{https://doi.org/10.1016/0550-3213(88)90152-6}{\emph{Nucl. Phys. B}
  {\bfseries 310} (1988) 335}.

\bibitem{Chemissany:2007he}
W.~Chemissany, M.~de~Roo and S.~Panda, \emph{{alpha'-Corrections to Heterotic
  Superstring Effective Action Revisited}},
  \href{https://doi.org/10.1088/1126-6708/2007/08/037}{\emph{JHEP} {\bfseries
  08} (2007) 037} [\href{https://arxiv.org/abs/0706.3636}{{\ttfamily
  0706.3636}}].

\bibitem{Grisaru:1986dk}
M.~T. Grisaru, A.~E.~M. van~de Ven and D.~Zanon, \emph{{Two-Dimensional
  Supersymmetric Sigma Models on Ricci Flat Kahler Manifolds Are Not Finite}},
  \href{https://doi.org/10.1016/0550-3213(86)90448-7}{\emph{Nucl. Phys. B}
  {\bfseries 277} (1986) 388}.

\bibitem{Gross:1986iv}
D.~J. Gross and E.~Witten, \emph{{Superstring Modifications of Einstein's
  Equations}}, \href{https://doi.org/10.1016/0550-3213(86)90429-3}{\emph{Nucl.
  Phys. B} {\bfseries 277} (1986) 1}.

\bibitem{Melnikov:2014ywa}
I.~V. Melnikov, R.~Minasian and S.~Sethi, \emph{{Heterotic fluxes and
  supersymmetry}}, \href{https://doi.org/10.1007/JHEP06(2014)174}{\emph{JHEP}
  {\bfseries 06} (2014) 174} [\href{https://arxiv.org/abs/1403.4298}{{\ttfamily
  1403.4298}}].

\bibitem{Candelas:1990rm}
P.~Candelas, X.~C. De~La~Ossa, P.~S. Green and L.~Parkes, \emph{{A pair of
  Calabi-Yau manifolds as an exactly soluble superconformal theory}},
  \href{https://doi.org/10.1016/0550-3213(91)90292-6}{\emph{Nucl. Phys.}
  {\bfseries B359} (1991) 21}.

\bibitem{Candelas:2018lib}
P.~Candelas, X.~De~La~Ossa, J.~McOrist and R.~Sisca, \emph{{The Universal
  Geometry of Heterotic Vacua}},
  \href{https://doi.org/10.1007/JHEP02(2019)038}{\emph{JHEP} {\bfseries 02}
  (2019) 038} [\href{https://arxiv.org/abs/1810.00879}{{\ttfamily
  1810.00879}}].

\bibitem{McOrist:2019mxh}
J.~McOrist and R.~Sisca, \emph{{Small gauge transformations and universal
  geometry in heterotic theories}},
  \href{https://doi.org/10.3842/SIGMA.2020.126}{\emph{SIGMA} {\bfseries 16}
  (2020) 126} [\href{https://arxiv.org/abs/1904.07578}{{\ttfamily
  1904.07578}}].

\bibitem{LopesCardoso:2003dvb}
G.~Lopes~Cardoso, G.~Curio, G.~Dall'Agata and D.~Lust, \emph{{BPS action and
  superpotential for heterotic string compactifications with fluxes}},
  \href{https://doi.org/10.1088/1126-6708/2003/10/004}{\emph{JHEP} {\bfseries
  10} (2003) 004} [\href{https://arxiv.org/abs/hep-th/0306088}{{\ttfamily
  hep-th/0306088}}].

\bibitem{delaOssa:2015maa}
X.~de~la Ossa, E.~Hardy and E.~E. Svanes, \emph{{The Heterotic Superpotential
  and Moduli}}, \href{https://doi.org/10.1007/JHEP01(2016)049}{\emph{JHEP}
  {\bfseries 01} (2016) 049}
  [\href{https://arxiv.org/abs/1509.08724}{{\ttfamily 1509.08724}}].

\bibitem{Ashmore:2018ybe}
A.~Ashmore, X.~De~La~Ossa, R.~Minasian, C.~Strickland-Constable and E.~E.
  Svanes, \emph{{Finite deformations from a heterotic superpotential:
  holomorphic Chern-Simons and an $L_\infty$ algebra}},
  \href{https://doi.org/10.1007/JHEP10(2018)179}{\emph{JHEP} {\bfseries 10}
  (2018) 179} [\href{https://arxiv.org/abs/1806.08367}{{\ttfamily
  1806.08367}}].

\bibitem{McOrist:2016cfl}
J.~McOrist, \emph{{On the Effective Field Theory of Heterotic Vacua}},
  \href{https://doi.org/10.1007/s11005-017-1025-0}{\emph{Lett. Math. Phys.}
  {\bfseries 108} (2018) 1031}
  [\href{https://arxiv.org/abs/1606.05221}{{\ttfamily 1606.05221}}].

\bibitem{McOrist:2025sdy}
J.~McOrist, M.~Sticka and E.~E. Svanes, \emph{{The heterotic G$_{2}$ moduli
  space metric}}, \href{https://doi.org/10.1007/JHEP11(2025)016}{\emph{JHEP}
  {\bfseries 11} (2025) 016}
  [\href{https://arxiv.org/abs/2502.16093}{{\ttfamily 2502.16093}}].

\bibitem{McOrist:2026cys}
J.~McOrist and Q.~Yin, \emph{{Heterotic moduli, the double extension and the
  $\alpha'^2$ metric}},
  \href{https://arxiv.org/abs/2607.15817}{{\ttfamily 2607.15817}}.

\bibitem{McOrist:2024glz}
J.~McOrist, M.~Sticka and E.~E. Svanes, \emph{{The moduli of the universal
  geometry of heterotic moduli}},
  \href{https://arxiv.org/abs/2411.05350}{{\ttfamily 2411.05350}}.

\end{thebibliography}

\providecommand{\href}[2]{#2}\begingroup\raggedright\endgroup

\end{document}